\documentclass[preprint,superscriptaddress,nofootinbib]{revtex4}

\usepackage{graphicx}

\begin{document}

 \title{Accelerated expansion in a stochastic self-similar fractal Universe}
\author{Eduardo Sergio Santini}
\affiliation{Centro Brasileiro de Pesquisas F\'{\i}sicas-MCT\\
     Lafex,
    Rua Dr.\ Xavier Sigaud 150, Urca 22290-180, Rio de Janeiro, RJ , Brasil}
\affiliation{ Comiss\~ao Nacional de Energia Nuclear-MCT \\ Rua General Severiano 90, Botafogo 22290-901 , Rio de Janeiro, RJ , Brasil}
\email{santini@cbpf.br} 
    \author{Guillermo Andr\'es Lemarchand}
 \affiliation{Facultad de Ciencias Exactas y Naturales, Universidad de Buenos Aires 
\\C.C. 8 - Sucursal 25, C1425FFJ Buenos Aires, Argentina }
\affiliation{Instituto Argentino de Radioastron\'{\i}a, Conicet, Argentina}
\email{lemar@correo.uba.ar}
\vspace{3cm}

\maketitle
\clearpage

\centerline{\bf Abstract}
\vspace{1.4cm}
    In a recent paper, a cosmological model based on El Naschie {\it E} infinity cantorian spacetime  was presented \cite{iovane}. In that work it was claimed that the present accelerated expansion of the Universe can be obtained as the effect of a scaling law on newtonian cosmology with a certain time dependent gravitational constant ($G$). In the present work we show that it may be problematic to explain the present accelerated expansion of the Universe using the approach presented in \cite{iovane}. As a better alternative  we apply the same scaling law and  a  time-dependent gravitational constant, that follows from the observational constraints, to relativistic cosmology, i.e. the Friedmann's model. We are able to show that for a matter-dominated flat  Universe, with the scaling law and a varying $G$, an accelerated expansion emerges in such a way that the function luminosity distance vs redshift can be made close to the corresponding  function that comes from  the usual Friedmann's model supplemented with a cosmological constant of value $\Omega_{\Lambda}\simeq0.7$. Then the measurements of high redshift supernovae, could be interpreted as a consequence of the fractal-self similarity of the $G$ varying relativistic universe.

\clearpage
 
    \section{Introduction}
    In recent years, several models were introduced to explain the measurements of high redshift supernovae \cite{SN1,SN} indicating that the Universe is presently in accelerated expansion. Two basic scenarios were discussed in the literature in order to take into 
account this observational result. The first one postulates, in the context of General Relativity, the existence of some negative pressure dark energy that, by opposing the self-atraction of matter, is causing the expansion of the universe to be positively accelerated \cite{triangle}\cite{garnavich}\cite{wang}. Candidates to be such dark energy are, among others, i) a cosmological constant, $\Lambda$, ii) a scalar field known as quintessence and iii) a Chaplygin gas \cite{martin}. The second scenario modifies General Relativity but without postulating the existence of a new contribution to the energy density as dark energy. Belonging to the last type of scenarios, several proposals were introduced such as (a) a quantum semiclassical cosmology approach \cite{parker}, (b) a quantum cosmology approach \cite{sergio} and (c) the recent Cantorian $\epsilon^{ \infty}$ approach \cite{iovane}. In the present work, we consider the cantorian approach model studied in \cite{iovane} and we analyse the validity of the Iovane's newtonian universe model  and the relativistic universe model. For the first model,  \cite{iovane}, we show that the deceleration parameter today is positive and then the expansion is decelerating. For the second model, we are able to show that it is possible to adjust the deceleration parameter today in order to be  negative with value $q_{0}\simeq -0.5$.This is compatible with a dark energy flat model with a cosmological constant ratio\footnote{$\Omega_{\Lambda} \equiv \frac{\rho_{vac}}{\rho^{0}_{crit}}=\frac{\Lambda}{3H_{0}^2}$ where $\rho^{0}_{crit}\equiv \frac{3H_{0}^2}{8\pi G}$
  is the present critical energy density of the universe and $\rho_{vac}$ is the vacuum energy density.} $\Omega_{\Lambda}\approx 0.7$.  Furthermore, we show that the luminosity distance for the studied model can be adjusted in close agreement with the corresponding one to the same dark energy flat model. These results suggest that it is possible to explain the accelerating expansion as an effect of a time dependent gravitational constant and the  Cantorian $\epsilon^{\infty}$ structure of the universe.

\section{The Iovane's  scenario}
Because luminous matter appears segregated at different scale as globular clusters, 
galaxies, clusters and superclusters of galaxies, Iovane  assumed in \cite{iovane} a scaling universal law of the form

\begin{equation}\label{sl}
R(N)=\lambda N^{\gamma}
\end{equation}
where $R$ is the radius of the astrophysical structure, $N$ is the number of nucleons in the 
structure and $\lambda$ is the Compton wavelength associated to a nucleon of mass
 $m_{n}$, $\lambda\equiv\frac{h}{m_{n}c}$, being $c$ the speed of light. This type of scaling law was also explored with different approaches by several authors in different large-scale universe contexts \cite{labini}\cite{WU}. The exponent $\gamma$ of the scaling 
law was derived from the theoretical framework developed by El Naschie in several publications (e.g. \cite{ELN}\cite{ELN1})  in a closed agreement to the Golden Section ($\phi \equiv \frac{\sqrt{5}-1}{2}$). Iovane et al. \cite{Iovane2}, using a stochastic self-similar and fractal universe model, estimated the value of $\gamma$ as 0.5.
In \cite{iovane}  the virial theorem and a scaling law are applied to a large-scale astrophysical structure to obtain a time-dependent gravitational constant, $G(t,N)$ given by\footnote{It was supposed in \cite{iovane}, apparently , that the product $\sqrt{GN}$ is constant in a fractal spacetime. However it must be noted that it is different in a fractal spacetime given by a fluctuational cosmological model, where $G\sqrt{N} = constant$, as was shown in \cite{sidharth}.}

\begin{equation}\label{gt}
G(t,N)=\overline{G} N^{3\gamma-1} t^{-2}
\end{equation}
where $\overline{G}=(\frac{4}{9})\frac{\lambda^3}{m_{n}} $.

\subsection{Hubble's law and the accelerated classical universe}

In the paper \cite{iovane} the time-dependence of the  Hubble's parameter was estimated by considering an object with velocity $v$ posed on the surface of a spherical mass with radius R. The expression for $v$ coming from the virial theorem,
 $v=\sqrt{\frac{GM}{R}}$ was substituted in the Hubble's law $v=HR$, being $H\equiv\frac{\dot{R}}{R}$ the Hubble's parameter. From here we can easily obtain the following relation

\begin{equation}\label{h}
H=\sqrt{\frac{GM}{R^3}}
\end{equation}
a different expresion to the one presented in Eq.(24) of \cite{iovane}, which is incorrect and dimensionally wrong.

If we substitute $R$ according to the scaling law (\ref{sl})  and $G$ according to (\ref{gt}) in Eq.(\ref{h}) 
the $\gamma$-dependence is cancelled and we obtain the following value for the Hubble constant:

\begin{equation}
H=\frac{2}{3}t^{-1}
\end{equation}

The deceleration parameter in $t_0\equiv$ today is

\begin{equation}
q_0 \equiv-(1+\frac{\dot{H}}{H^2})\mid_0
\end{equation}
which results in a value of

\begin{equation}
q_0 =\frac{1}{2}
\end{equation}
and then the expansion today is decelerated. 
We  see that no accelerated expansion is predicted by the Iovane's approach to a Newtonian cosmology,   as claimed in \cite{iovane} and other recent papers \cite{Iovane2} \cite {IO1} \cite {IO2} \cite {IO3} \cite {IO4}.

However, following the Iovane's ideas and going one step further, we will consider the consequences that,  
the scaling law and a time dependence of $G$, have in a relativistic universe, for instance, a dust matter Friedmann flat model.\footnote{We know that it is not strictly a Friedmann model because the time dependence of $G$. 
We are aware of the limitations of this consideration because we are in fact  working with a "beyond standard Friedmann model", and this is to a certain matter a speculation.
It will be interesting to consider a Brans-Dicke cosmology, in which the gravitational constant is varying with $t$ as a field [20][21]. This will be the subject of our future investigations.
} \footnote{ A relativistic non-homogeneous model, the Tolman-Bondi model, in a cantorian spacetime was already  studied in \cite{IO3}.} It is posible to obtain an accelerated expansion as an effect of the scaling law and a certain time-dependence of $G$, i.e.  the Cantorian structure of spacetime, as we will see in the following sections. 

\section{Scaling law and time varying $G$ relativistic model}

The Friedmann's equation for a  matter-dominated universe with flat spatial section is

\begin{equation}\label{Fried0}
H^2=\frac{8\pi G}{3}\rho
\end{equation}
where $\rho$ is the energy density of matter (we consider $c=1$). Since stable 
matter is not spontaneously created or destroyed, we have $\rho=\rho_0 \frac{R_{0}^{3}}{R^3}$, where $R(t)$ is the scale factor of the Universe and the subscript $0$ stands for the present time. Then

\begin{equation}\label{Fried}
H^2=\frac{8\pi G}{3}\rho_0 \frac{R_{0}^{3}}{R^3}.
\end{equation}

We will assume a time dependence for $G$ which follows from the observational constraints given in the literature as an upper limit on $\mid \dot{G}/G\mid$:
\begin{equation}\label{Gexp}
\mid\dot{G}/G \mid < B.
\end{equation}
where $10^{-13}yr^{-1} \leq  B \leq 10^{-10}yr^{-1}$ \cite{uzan}.
Then, we write

\begin{equation}\label{Gexp=}
\dot{G}/G = -b.
\end{equation}
for $b<B$

In fact, by integrating Eq. (\ref{Gexp=}) we get 
\begin{equation}\label{Gexp2}
G=G(t_0) e^{-b(t-t_0)}.
\end{equation}

Now, following the Iovane's approach, we substitute  in Eq. (\ref{Fried}) $R$ according to  the scaling law (Eq. (\ref{sl})) and $G$  according to Eq. (\ref{Gexp2}). Considering that  
$\rho_0 Volume=\rho_0 v R_{0}^{3}=N m_n$ where $v$ is a constant, we have
\footnote{We recall that $v$ is equal to the volume divided by $R^3$ of the spatial sections, which are  supposed to 
be closed. The quantity $v$ depends on the spatial curvature and on the topology. 
For the case of flat spatial section it  can have any value because the fundamental polyhedra with closed flat hypersurfaces can have arbitrary size \cite{vitorio}. For the case of positive curvature and topology $S^3$ we have $v=2\pi^2$.}

\begin{equation}\label{Fried2}
H=H_0 e^{\frac{b}{2}t_0}e^{-\frac{b}{2}t}
\end{equation}
where

\begin{equation}\label{Fried3}
H_0=\sqrt{\frac{8\pi G(t_0)}{3 h^3 }\frac{(m_n c)^3 m_n}{N^{3\gamma-1}v}}
\end{equation}

Taking $H_0=6.7 \times 10^{-11}yr^{-1}=2.1\times10^{-18}s^{-1} $(i.e $67km s^{-1} Mpc^{-1}$) \cite{peebles}, $G(t_0)=6.67 \times 10^{-11}m^{3}Kg^{-1}s^{-2}$, $h=6.626 \times 10^{-34}j s$, $N\simeq10^{80}$, $m_{n}=1.673\times10^{-27}Kg$, recovering the velocity of light as $c = 3 \times  10^{8}m s^{-1}$, and assuming that $v=\frac{4\pi}{3}$, Eq. (\ref{Fried3}) gives for the scaling power $\gamma \simeq 0.51$ (For $v=10^{(-23)}$ we have $\gamma\simeq\phi=\frac{\sqrt{5}-1}{2}$).
Note that from Eq. (\ref{Fried3}) the value of the scaling $\gamma$ is related to the spatial topology 
through the quantity $v$.

The deceleration parameter today results as 
\begin{equation}\label{q}
q_0 \equiv-(1+\frac{\dot{H}}{H^2})\mid_0
= -1+\frac{b}{2H_0} 
\end{equation}
which is negative if $\frac{b}{2H_0}<1$, indicating an accelerated expansion.

This is the case if we assume $b \simeq H_0\simeq 6.7 \times 10^{-11}yr^{-1}$, obtaining 

\begin{equation}\label{q05}
q_0\simeq -0.5
\end{equation}

Note that $b$ could take even  a lower value, according to the observational constraint given by  Eq.(\ref{Gexp}). Then, according to Eq.(\ref{q}), the deceleration parameter should be more negative, i.e. the expansion is even more accelerated.

\subsection{Comparison between the accelerated expansion of the fractal Universe and the presence of a cosmological constant}
\subsubsection{The deceleration parameter}
Now we will compare this model with a flat Friedmann universe with dust matter and a cosmological constant  in such a way that it is experiencing an accelerated expansion.

We recall that  a flat model  universe containing matter and a cosmological constant $\Lambda$ (equivalently, vacuum energy)
 with ratios values $\Omega_m$ and $\Omega_{\Lambda}$ respectively\footnote{It is ussualy defined $\Omega_m\equiv \frac{\rho_0}{\rho^{0}_{crit}}=\frac{8\pi G\rho}{3H_{0}^2}$ 
and, as already defined, $\Omega_{\Lambda} \equiv \frac{\rho_{vac}}{\rho^{0}_{crit}}=\frac{\Lambda}{3H_{0}^2}$ 
where $\rho^{0}_{crit}\equiv \frac{3H_{0}^2}{8\pi G}$
  is the present critical energy density of the universe and $\rho_{vac}$ is the vacuum energy density.}, 
satisfies the Friedmann's equation 

\begin{equation}
H^2=\frac{8\pi G}{3}\rho_0 \frac{R_{0}^{3}}{R^3}+\frac{\Lambda}{3}=H_0^2[(1-\Omega_\Lambda)(1+z)^3+\Omega_\Lambda ]
\end{equation}
where $1+z\equiv\frac{R_0}{R}$ indicates the redshift.

The deceleration parameter is 

\begin{equation}\label{ql}
q_0^{(\Lambda)}=\frac{\Omega_m}{2}-\Omega_\Lambda = \frac{1}{2}-\frac{3}{2}\Omega_\Lambda 
\end{equation}
where for a flat universe we take $\Omega_m + \Omega_\Lambda = 1$. If we use $\Omega_\Lambda=0.7$ in 
Eq. (\ref{ql}) we obtain $q_0^{(\Lambda)}=-0.55$, i.e. approximately the value obtained in Eq.(\ref{q05}).Then the cantorian model studied is compatible with a cosmological constant with that value. 
 The Wilkinson Microwave Anisotropy Probe (WMAP) together with the Sloan Digital Sky Survey give $\Omega_\Lambda\approx 0.7 $ \cite{tegmark, tegmark1}. 

\subsubsection{Luminosities distances}

The supernovae measurements relate the luminosity distance $d$ with the redshift $z$ in the following way:

\begin{equation}\label{dl}
d=(1+z)\int_{0}^{z}\frac{{\rm d}y}{H(y)}.
\end{equation}

For the model considered with cosmological constant, calling the luminosity distance $d^{\Lambda}$, we have

\begin{equation}\label{dlc}
d^{\Lambda}=\frac{(1+z)}{H_0}\int_{0}^{z}\frac{{\rm d}y}{\sqrt{(1-\Omega_\Lambda)(1+y)^3+\Omega_\Lambda }}.
\end{equation}

We integrate Eq. (\ref{Fried2}) for the studied cantorian model, and using the definition of Hubble parameter, 
we obtain, for the scale factor the following relation

\begin{equation}\label{scale}
R(t)=R_0 e^{-\frac{2}{b}H_0 (e^{-\frac{b}{2}(t-t_o)}-1)}
\end{equation}
then we can rewrite Eq.(\ref{Fried2}) as

\begin{equation}\label{Fried4}
H=H_0 (1+ \frac{b}{2H_0}\ln{(1+z)}).
\end{equation}

Eq (\ref{Fried4}) gives the particular $z$-dependence of the Hubble parameter coming from the scaling law-varying $G$ (i.e. cantorian) structure of the relativistic model studied here.
 We want to know what is the luminosity distance $d^{c}$ that this particular "fractal associated" behaviour of $H(z)$ determines, according to the definition (\ref{dl}).   Now, substituting in Eq. (\ref{dl}) the expression for $H(z)$ given by Eq. (\ref{Fried4}) , we have for the luminosity distance\footnote{As a first rough approximation we calculate this integral in the Riemann measure although this  can be a matter of discussion, because the Hausdorff measure is the more appropriate to a fractal structure. However a fractal contribution to the luminosity distance, that we call $d^{c}$, is already given by the particular behaviour of $H(y)$ in the Riemann integral, Eq. (\ref{dlum}).}:

\begin{equation}\label{dlum}
d^{c}=\frac{(1+z)}{H_0}\int_{0}^{z}\frac{{\rm d}y}{(1+ \frac{b}{2H_0}\ln{(1+y)})}.
\end{equation}
In Fig. 1 we plot the luminosities distances (times $H_0$) $H_0 d^{c}(z)$, given by Eq. (\ref{dlum}) for $b=1.8H_0$,  and $H_0 d^{\Lambda}(z)$, given by Eq. (\ref{dlc}), for $\Omega_\Lambda=0.7$. For small and intermediate values of $z$, the curve corresponding to the cantorian relativistic model, is close to the one corresponding to the Friedmann flat model with matter and a cosmological constant. These $z$ values are consistent with recently discovered highest redshift SNe Ia, all at $z>1.25$ \cite {RIESS}. For higher $z$ both models separate. 

These results suggest, that a fractal self-similar structure universe  could emulate the effect of a  cosmological constant, or in other words, that a connection between the accelerated expansion today and  the Cantorian $\epsilon^{ \infty}$ structure of spacetime may exist. Then the present observations of high redshift supernovae could be interpreted as the manifestation of the fractal self-similarity of the universe.

\begin{figure}
\includegraphics{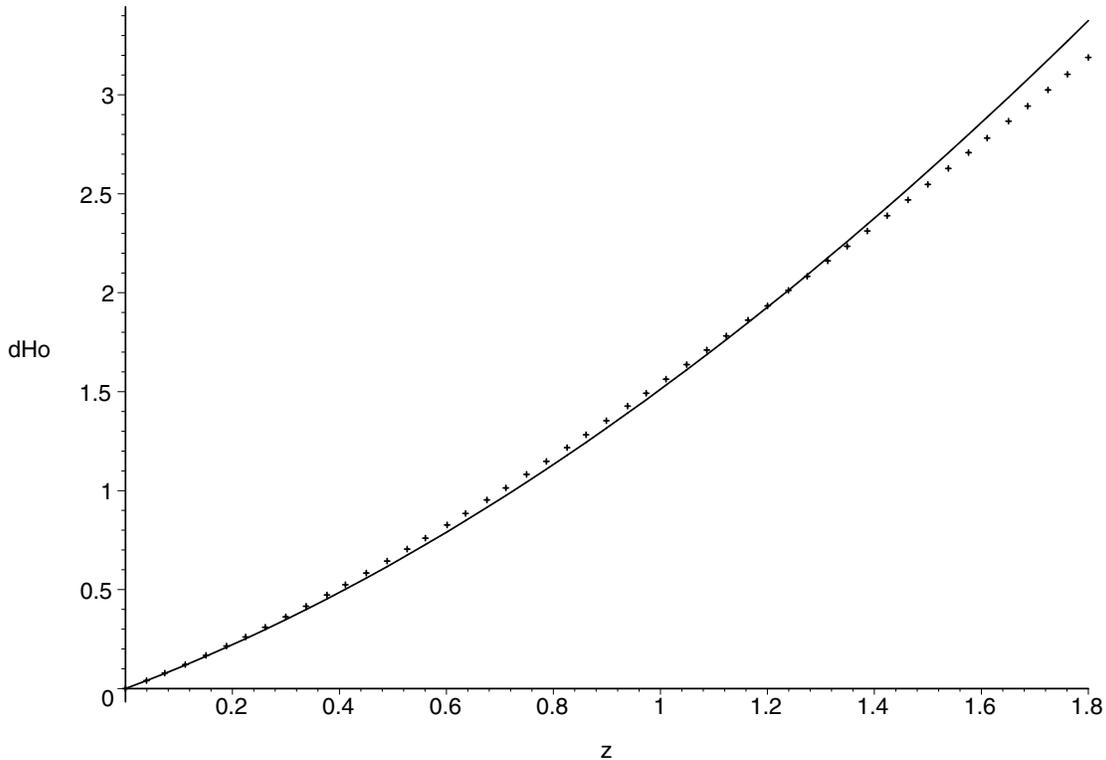}
\caption{Luminosities distances as a function of the redshift. The dotted curve corresponds to the cantorian model with $b=1.8 H_0$ and the line curve to the Friedmann dust matter flat model with a cosmological constant $\Omega_{\Lambda}=0.7$.}
\label{cantor}
\end{figure}

\section{Conclusions}
We have studied the possible connection between the accelerated expansion of the universe and 
its  fractality. Following an idea of Iovane we explored the consequences of 
a scaling law and a time-dependent gravitational constant on cosmological models. In the case of 
the Iovane's newtonian model, we show that there is no accelerated expansion  \cite{iovane}.
For the case of a relativistic cosmological model, given by a flat dust matter Friedmann's model, we were able to show that the corresponding deceleration parameter today  can be adjusted to be approximately  equal to the one corresponding  to a (dark energy) model with matter plus a cosmological constant with  value $\Omega_\Lambda\approx 0.7$.  This result is compatible with the  the most accurate measurement accepted today (i.e $\Omega_\Lambda\approx 0.7$). Furthermore, we have shown that it is possible to adjust the model in order to obtain a luminosity distance as a function of the redshift in a close agreement with the luminosity distance of that (dark energy) model.
In such a way, the present observations of high redshift supernovae could be interpreted, not as the existence of 
a dark energy contribution, but as a consequence of a scaling law (i.e. fractality) and a time-varying gravitational constant operating in the universe. We recall that we have only studied a toy model.  More elaborated models, taking into account other relevant sources like radiation, and exploring different $t$-dependence of $G$  already discussed in the literature (i.e. Brans-Dicke theory) must be explored. This will be the subject of our future investigations.

\section{Acknowledgements}
ESS want to thank Minist\'erio da Ci\^encia e Tecnologia (MCT-CBPF and CNEN) of Brazil and International Atomic Energy Agency for finnancial support. GAL was supported by the UBACyT U401 Research Project. We want to thank to L. Nottale for useful references. We thank to the Referee for valuable critical comments and suggestions.

\end{document}